\documentclass[prl,aps,twocolumn,preprintnumbers, showpacs, nofootinbib,superscriptaddress,notitlepage]{revtex4-1}
\usepackage{mathrsfs}
\usepackage{color}
\usepackage{enumitem}
\usepackage{amsfonts}
\usepackage{amsmath}
\usepackage{slashed}
\usepackage{array}
\usepackage{verbatim}
\usepackage{epsfig}
\usepackage{graphicx}
\usepackage{color}

\newcommand{\beq}{\begin{eqnarray}}
\newcommand{\eeq}{\end{eqnarray}}

\def \non {\nonumber}

\begin{document}

\preprint{}

\title{Accessing gluon parton distributions in large momentum effective theory}

\author{Jian-Hui Zhang}
\affiliation{Institut f\"ur Theoretische Physik, Universit\"at Regensburg, D-93040 Regensburg, Germany}

\author{Xiangdong Ji}
\affiliation{Maryland Center for Fundamental Physics,
Department of Physics, University of Maryland,
College Park, Maryland 20742, USA}
\affiliation{Tsung-Dao Lee Institute, and School of Physics and Astronomy,
Shanghai Jiao Tong University, Shanghai, 200240, China}

\author{Andreas Sch\"afer}
\affiliation{Institut f\"ur Theoretische Physik, Universit\"at Regensburg, D-93040 Regensburg, Germany}

\author{Wei Wang}
\email{wei.wang@sjtu.edu.cn}
\affiliation{SKLPPC, MOE KLPPC, School of Physics and Astronomy, Shanghai Jiao Tong University, 800 Dongchuan RD, Shanghai, 200240, China}

\author{Shuai Zhao}
\affiliation{SKLPPC, MOE KLPPC, School of Physics and Astronomy, Shanghai Jiao Tong University, 800 Dongchuan RD, Shanghai, 200240, China}

\date{\today}

\begin{abstract}
Gluon parton distribution functions (PDFs) in the proton can be calculated directly on Euclidean lattices using large 
momentum effective theory (LaMET). To realize this goal, one has to find renormalized gluon quasi-PDFs in which 
power divergences and operator mixing are thoroughly understood. For the unpolarized distribution, 
we identify four independent quasi-PDF correlators that can be multiplicatively renormalized
on the lattice. Similarly, the helicity distribution  can be derived from three independent multiplicatively renormalizable 
quasi-PDFs.  We provide a LaMET factorization formula for these renormalized quasi-PDFs from which one can extract the gluon PDFs.
\end{abstract}
\maketitle

{\bf Introduction:}
Parton distribution functions (PDFs) are crucial quantities characterizing the structure of hadrons at high energy. They are defined as momentum distributions of quarks and gluons in an infinite-momentum hadron, and provide a necessary input for the description of experimental data from hadron-hadron or lepton-hadron colliders. However, calculating   PDFs from first principles has been a long-standing problem in hadron physics, since PDFs are low-energy properties of a  hadron defined in terms of lightcone correlations. Traditional lattice QCD methods   focus  on the calculation of their moments, but can only determine the first few moments   due to the power divergent mixing between different moment operators~\cite{Martinelli:1987zd,Martinelli:1988xs,Detmold:2001dv,Dolgov:2002zm}. Reconstructing   PDFs, however, requires information on all their moments.

In the past few years, a new approach has been proposed to circumvent the above difficulty, which has now been formulated as the large momentum effective theory (LaMET)~\cite{Ji:2013dva,Ji:2014gla}. LaMET provides a possibility to study parton properties of the hadron in general. For a given parton observable such as   PDFs, one can deconstruct the lightcone correlations such that the resulting quantity, known as a quasi-observable, becomes time-independent though no longer frame-independent. The quasi-observable approaches the original parton observable under an infinite Lorentz boost, and has the same infrared (IR) behavior as the latter by construction. In a hadron with finite but large momentum, the quasi-observable can be factorized into the parton observable convoluted with a hard coefficient, up to power corrections suppressed by the hadron momentum~\cite{Ji:2013dva,Ji:2014gla,Xiong:2013bka,Ma:2017pxb,Izubuchi:2018srq,Ji:2015qla,Xiong:2015nua}.
Since LaMET was proposed, much progress has been achieved with respect to both the theoretical understanding of the formalism and the direct calculation of PDFs from lattice QCD (see a recent review~\cite{Cichy:2018mum} and references therein). The lattice calculations have been predominantly focused on isovector quark PDFs, since they do not mix with gluon PDFs. Despite limited volumes and relatively coarse lattice spacings, the state-of-the-art nucleon isovector quark PDFs determined from lattice data at the physical point already show a reasonable agreement~\cite{Chen:2018xof,Lin:2018qky,Alexandrou:2018pbm,Alexandrou:2018eet,Liu:2018hxv} with phenomenological results extracted from experimental data~\cite{Dulat:2015mca,Ball:2017nwa,Harland-Lang:2014zoa,Nocera:2014gqa,Ethier:2017zbq}.

Lattice calculations of gluon parton physics are even more crucial because gluons play an extremely important role in experiments performed at the Large Hadron Collider (LHC) and the future Electron-Ion Collider (EIC) in US. However,  
in contrast to the intensive work on extracting quark PDFs using LaMET, much less effort has been devoted to the study of gluon quasi-PDFs.  
In Refs.~\cite{Wang:2017qyg,Wang:2017eel}, the first effort was made in understanding the renormalization property of gluon quasi-PDFs, where it was discovered that the gluon quasi-PDF suffers from a mixing with other gluon operators allowed by symmetries of the theory. While on the other hand,  in the first exploratory attempt to calculate the gluon unpolarized PDF~\cite{Fan:2018dxu}, the authors assumed a multiplicative renormalization of the gluon quasi-PDF used in their lattice calculation.

In this Letter, we perform a systematic study of the renormalization of gauge-invariant non-local operators defining the gluon quasi-PDFs, focusing on their
power divergence structure. Following an earlier work involving two of us~\cite{Ji:2017oey} (see also~\cite{Ishikawa:2017faj,Green:2017xeu}), we introduce   a general framework in which the Wilson line in the adjoint representation is replaced by a product of two   auxiliary ``heavy quark" fields. Such a theory with an auxiliary   ``heavy quark"  can be shown to be renormalizable to all orders in perturbation theory.  
We  then show how to construct a multiplicatively renormalizable gluon quasi-PDF operator, and explain how to calculate the renormalization factors non-perturbatively on lattice. The analysis is also extended to the polarized case. At last, we present a factorization formula for the renormalized gluon quasi-PDF, from which one can extract the gluon PDF. Our findings remove the obstacle to define a continuous gluon quasi-PDF through lattice simulations.

{\bf Auxiliary adjoint ``heavy quark" approach:}
Let us start with the gluon quasi-PDF defined in Ref.~\cite{Ji:2013dva}
\begin{align}\label{gluonpdfop}
\tilde g(x, P^z)=\int\frac{dz}{2\pi x n\cdot P}e^{i x z n\cdot P}\langle P|O(z,0)|P\rangle,\non\\
O(z_2,z_1)=F_a^{z\mu}(z_2)L_{ab}(z_2, z_1)F_{b,\mu}^z(z_1).
\end{align}
$F_a^{\mu\nu}=\partial^\mu A_a^{\nu}-\partial^\nu A_a^\mu-g f_{abc}A_b^\mu A_c^\nu$ is the field strength tensor. Unless otherwise stated, the quantities defined in this section are all bare ones. The index $\mu$ can be summed either over transverse components, as in the gluon PDF~\cite{Collins:1984xc}; or over all Lorentz components. Since $F^{zz}=0$ and $F^{zt}$ gets power suppressed in the large momentum limit, both summations can be used to define the gluon quasi-PDF and they fall into a universality class of the gluon quasi-PDF operators~(for discussion on universality class of the gluon polarization operator, see Ref.~\cite{Hatta:2013gta}).  $P^\mu=(P^0, 0,0, P^z)$ is the momentum of the external hadron, and $n^\mu=(0,0,0,1)$ is a spacelike vector specifying the direction of the gauge link $L(z, 0) = {\cal P}\exp\left(-ig\int^z_0 d\lambda\,n^\mu  A_\mu(\lambda n)\right)$ with $A_\mu=A_{\mu, a}T_a$ being defined in the adjoint representation.

Alternatively, the gluon quasi-PDF can be defined using fundamental gauge links $U(z_2,z_1)$~\cite{Dorn:1981wa}
\beq\label{fundrep}
\hspace*{-1em}O(z_2,z_1)=2\,{\rm Tr}[F^{z\mu}(z_2)U(z_2,z_1)F_{z\mu}(z_1)U(z_1,z_2)],
\eeq
where $F^{\mu\nu}=F_{a}^{\mu\nu}t_a$ and $t_a$ is the generator in the fundamental representation.
While Eq.~(\ref{gluonpdfop}) facilitates the study of renormalization, Eq.~(\ref{fundrep}) is more straightforward for lattice implementations. We will mainly focus on the definition in Eq.~(\ref{gluonpdfop}) below, but the results also apply to Eq.~({\ref{fundrep}}).

To renormalize $O(z_2,z_1)$, we introduce an auxiliary adjoint ``heavy quark" field, denoted as $\mathcal Q$, into the QCD Lagrangian, following our study of the quark quasi-PDF~\cite{Ji:2017oey}. This ``heavy quark" has trivial spin degrees of freedom.  The Lagrangian with the auxiliary field reads
\beq\label{efflag}
\mathcal L=\mathcal L_{\rm{QCD}}+\overline{\mathcal Q}(x)(i n\cdot D-m)\mathcal Q(x),
\eeq
where $D_\mu=\partial_\mu+ig A_{\mu}$ is the covariant derivative in the adjoint representation, and a ``residual mass"   is introduced for the auxiliary field $\mathcal Q$, for reasons that will become clear later.

With the auxiliary ``heavy quark" $\mathcal Q$,  the Wilson line can be replaced by a product of two such fields~\cite{Ji:2017oey}
\begin{align}\label{Eq:repl}
{\cal O}(z_2, z_1) &= F_a^{z\mu}(z_2){\mathcal Q}_a(z_2) \overline{\mathcal Q}_b(z_1) F^z_{b, \mu}(z_1)\non\\
&= J_1^{z\mu}(z_2) {\overline J}_{1, \mu}^z(z_1). 
\end{align}
The renormalization of the non-local operator in Eq.~(\ref{gluonpdfop}) now  follows from that of the product of two local composite operators (LCOs) in Eq.~(\ref{Eq:repl}).

{\bf Renormalization in auxiliary field approach:}
The renormalization of the effective Lagrangian Eq.~(\ref{efflag}) is in complete analogy to that with a fundamental auxiliary field~\cite{Bagan:1993zv,Ji:2017oey}, and suffices to render Green's functions of the elementary fields finite. However, extra complications arise when renormalizing the LCOs $J_1, {\overline J}_1$ in Eq.~(\ref{Eq:repl}). 
We first consider the perturbative renormalization in dimensional regularization (DR) in a covariant gauge. In such a case, gauge-invariant LCOs can, in general, mix with operators of the same or lower mass dimension under renormalization. The mixing operators can be of the following three types: 1) gauge-invariant operators, 2) BRST exact operators,
3) operators that vanish by equation of motion (EOM)~\cite{Collins:1984xc}. For the LCO formed by a quark and a fundamental auxiliary  field, such a mixing does not occur because the operator appearing there is already of the lowest dimension and, if chiral symmetry is preserved, the mixing between operators of the same dimension is forbidden~\cite{Chen:2017mie}. This is no longer true in the present case. 

In DR, the operators allowed by BRST symmetry to mix with $J_1^{\mu\nu}$ are
\begin{align}\label{mixingop}
J_2^{\mu\nu}&=n_{\rho}(F^{\mu\rho}_a n^\nu-F^{\nu\rho}_a n^{\mu}){\mathcal Q}_a/n^2,\non\\
J_3^{\mu\nu}&=(-in^\mu A_a^{\nu}+in^\nu A_a^\mu)((in\cdot D-m) \mathcal Q)_a/n^2,
\end{align}
where $J_2^{\mu\nu}$ is gauge invariant, $J_3^{\mu\nu}$ is proportional to the EOM of $\mathcal Q$ and therefore vanishes in a physical matrix element (for $m=0$ the above mixing pattern has been given in~\cite{Dorn:1981wa}).

In the presence of mixing, the renormalized operators can be written in a  triangular mixing form
\begin{align}\label{oprenormmix}
\begin{pmatrix}
 J_{1, R}^{\mu\nu} \\  J_{2, R}^{\mu\nu} \\ J_{3, R}^{\mu\nu}
\end{pmatrix}
&=
\begin{pmatrix}
Z_{11} & Z_{12} & Z_{13}\\ 0 & Z_{22} & Z_{23} \\ 0 & 0 & Z_{33}
\end{pmatrix}
\begin{pmatrix}
 J_1^{\mu\nu}\\  J_2^{\mu\nu} \\ J_3^{\mu\nu}
\end{pmatrix}.
\end{align}
The renormalization constants in Eq.~(\ref{oprenormmix}) are not all independent. When $\nu=z$, $J_2^{z\mu}$   becomes degenerate with $J_1^{z\mu}$,
therefore they must have the same renormalization, implying
\beq\label{Zrelation}
Z_{11}+Z_{12}=Z_{22}, \hspace{5em} Z_{13}=Z_{23}.
\eeq
This is consistent with explicit one-loop calculations~\cite{Dorn:1981wa}.
The above result indicates that the individual components of $J_1^{\mu\nu}$ renormalize independently with the following simplified renormalization equations:
\begin{align}\label{oprenormmixsim2x2}
\begin{pmatrix}
 J_{1, R}^{z\mu} \\ J_{3, R}^{z\mu}
\end{pmatrix}
&=
\begin{pmatrix}
Z_{22} & Z_{13} \\  0 & Z_{33}
\end{pmatrix}
\begin{pmatrix}
 J_1^{z\mu} \\ J_3^{z\mu}
\end{pmatrix};\hspace{1em}
J_{1, R}^{ti}=Z_{11} J_1^{ti}.
\end{align}
$J_{1}^{ij}$ shares the same renormalization with $J_{1}^{ti}$. The reason that
($J_{1}^{ti}$, $J_{1}^{ij}$) and $J_1^{z\mu}$ operators have different renormalizations is because the presence of a four-vector $n^\mu$ breaks Lorentz symmetry. Note that $J_2^{\mu i}$ is not independent, and henceforth  will be neglected.

Our calculation show that no linear divergences occur in the one-loop correction to $J_1^{\mu\nu}$. Actually, the power divergence structure of this operator has been considered in Refs.~\cite{Wang:2017qyg,Wang:2017eel}, but using a simple cutoff regularization. One has to be cautious when dealing with power divergences in cutoff regularization, since it in general breaks gauge invariance in QCD (except for the lattice cutoff), and might obscure the structure of genuine power divergences of the theory. To avoid this, we have chosen to work in DR and kept track of linear divergences by expanding around $d=3$, as they appear as poles at $d=3$. In this way, we can extract linear divergences gauge invariantly and confirm that no linear divergences occur in the one-loop correction to $J_1^{\mu\nu}$. The same conclusion can actually be reached in Ref.~\cite{Wang:2017eel} too, if a gauge-invariant regulator is used. We have explicitly checked this following the procedure outlined above.

In Eq.~(\ref{efflag}), we also introduced a ``residual mass" term for the auxiliary field $\mathcal Q$. The reason is to keep track of potential mass renormalization, which might be absent in DR when expanded at $d=4$,  but nevertheless appears in a cutoff regularization such as lattice regularization. In a cutoff regularization, an effective mass for the auxiliary ``heavy quark" will be generated by radiative corrections even if it does not exist at leading order. 
This is indeed what happens in our present case, and we have $m=\delta m\sim{\cal O}(1/a)$ with $a$ being the inverse cutoff or lattice spacing~\cite{Maiani:1991az}. In perturbation theory, $\delta m$ appears from $O(\alpha_s)$. 
Such a mass term serves the purpose of absorbing power divergences arising from the Wilson line self energy when integrating out the auxiliary field, as will be seen below.  Apart from this, there is no other power divergence in the theory. Moreover, for dimensional reasons, there is no other antisymmetric operator that can mix with $J_i^{\mu\nu}$ discussed previously. Therefore in a gauge-invariant cutoff scheme, the operator renormalization remains the same as in DR.

In Ref.~\cite{Ji:2017oey} it was shown that BRST invariance of the Lagrangian requires a dependence of $m$ on the signature of $n$~\cite{Dorn:1986dt}, which yields a vanishing mass for a lightlike $n^\mu$ and a real mass for a timelike $n^\mu$. When $n^\mu$ is spacelike, $m=\delta m=i\overline{\delta { m}}$ is imaginary. The value of $\overline{\delta { m}}$ at $O(\alpha_s)$ can be easily obtained from Ref.~\cite{Ji:2017oey} as $\overline{\delta { m}}={\pi\alpha_s C_A}/(2a)$.

{\bf Gluon quasi-PDF operators:}
Now we are ready to construct an appropriate gluon quasi-PDF operator.
To this end, we need to identify one of the indices in $J_1^{\mu\nu}$ with $z$ or $t$ and let the other run either over all Lorentz components or over transverse components only. It is worthwhile to point out at this stage that the  operator $J_3^{\mu\nu}$ only yields contact terms when integrating out the ``heavy quark" field. This can be seen from the fact that the EOM operator acting on the ``heavy quark" propagator yields a $\delta$-function. The contact terms do not vanish only when $z_2=z_1$, indicating that an extra renormalization is required when two LCOs shrink to one spacetime point. For the renormalization of the non-local gluon quasi-PDF operator, $J_3^{\mu\nu}$ is irrelevant and can be ignored.

From the discussions above, we identify the following building blocks, $J_{1,R}^{zi}$, $J_{1,R}^{ti}$, $J_{1,R}^{z\mu}$, for the gluon quasi-PDF operator. A multiplicatively renormalizable gluon unpolarized quasi-PDF operator can be constructed from
\begin{align}
 {\cal O}^{1}_R(z_2,z_1) &\equiv  J_{1,R}^{ti}(z_2) \overline J_{1,R}^{ti}(z_1), \non\\
 {\cal O}^{2}_R(z_2,z_1) &\equiv  J_{1,R}^{zi}(z_2)\overline J_{1,R}^{zi}(z_1), \nonumber\\
 {\cal O}^{3}_R(z_2, z_1) &\equiv    J_{1,R}^{ti}(z_2) \overline J_{1,R}^{zi}(z_1),\non\\
 {\cal O}^{4}_R(z_2, z_1)  &\equiv  J_{1,R}^{z\mu}(z_2) \overline J_{1,R,\mu}^{z}(z_1).
\end{align}
After integrating out the auxiliary ``heavy quark" field~\cite{Ji:2017oey}, these operators renormalize multiplicatively as
\begin{align}
{O}^{1}_R(z_2, z_1) &=Z_{11}^2 e^{\overline{\delta { m}}\Delta z}F^{ti}(z_2) L(z_2,z_1) F^{ti}(z_1),\nonumber\\
 {  O}^{2}_R(z_2,z_1) &=Z_{22}^2 e^{\overline{\delta { m}}\Delta z} F^{zi}(z_2) L(z_2,z_1) F^{zi}(z_1),\nonumber\\
 {O}^{3}_R(z_2,z_1) &= Z_{11}Z_{22} e^{\overline{\delta { m}}\Delta z} F^{ti}(z_2) L(z_2,z_1) F^{zi}(z_1),\nonumber\\
O^{4}_R(z_2,z_1)
& = Z_{22}^2 e^{\overline{\delta { m}}\Delta z} F^{z\mu}(z_2) L(z_2,z_1) F^{z}_{\;\;\mu}(z_1),
 \end{align}
with $\Delta z=|z_2-z_1|$.

In the large momentum limit,
the operators  $O^{i}_R(i=1,2,3,4)$ differ from each other only by power corrections. They therefore belong to the same universality class defining the gluon quasi-PDF.
Given the above four operators, one can use any combination  of them to study the gluon quasi-PDF, however such combinations are usually not multiplicatively renormalizable. A notable example is
\begin{align}
{  O}^{5}_R(z_2,z_1) 
 =    -  {  O}^{1}_R(z_2,z_1)-  {  O}^2_R(z_2, z_1)-{  O}^4_R(z_2, z_1),
\end{align}
which (minus the trace term) has been used in the recent simulation~\cite{Fan:2018dxu}.
However, since the renormalizations for ${  O}^{1}_R(z_2,z_1)$ and  $ {  O}_R^{2,4}(z_2, z_1)$ are different, ${  O}_R^{5}(z_2,z_1)$ is not multiplicatively renormalizable.

In the above discussion, we considered the gluon quasi-PDF only. If we insert the gluon quasi-PDF operator into a quark state, it   gives rise to UV finite contributions, provided that all subdivergences have been renormalized. The reason is that, the quasi-PDF is defined at spacelike separations, therefore there is no non-local UV divergence apart from the exponential mass renormalization. It indicates that the gluon quasi-PDF does not mix with the quark ones under renormalization in the above sense, while the mixing occurs at the factorization stage. This has been confirmed by the one-loop calculations in Ref.~\cite{Wang:2017qyg}, where  the quark matrix element of the gluon quasi-PDF operator is UV finite, but contains $\ln P_z^2/p^2$ ($p^2$ is an IR regulator) terms associated with the corresponding splitting kernel. This will be matched to the quark PDF to ensure the cancellation of IR divergences between the quasi-PDF and the PDF. 

{\bf Gluon helicity distribution:}
Following the same procedure, we can define three operators that can be used to calculate the gluon helicity distribution:
\begin{align}
\Delta {O}^{1}_R(z_2, z_1) &= Z_{11}^2 e^{\overline{\delta { m}}\Delta z}  \epsilon_{ij}  F^{ti}(z_2) L(z_2,z_1) F^{tj}(z_1), \non\\
 \Delta{  O}^{2}_R(z_2,z_1) &=  Z_{22}^2 e^{\overline{\delta { m}}\Delta z} \epsilon_{ij}  F^{zi}(z_2) L(z_2,z_1) F^{zj}(z_1),\non\\
\Delta {O}^{3}_R(z_2,z_1) &=  Z_{11}Z_{22} e^{\overline{\delta { m}}\Delta z}  \epsilon_{ij} F^{ti}(z_2) L(z_2,z_1) F^{zj}(z_1),
\end{align}
where $ \epsilon_{ij} $ is the two-dimensional anti-symmetric tensor.

{\bf Implementation on lattice:}
To define a renormalized gluon quasi-PDF on the lattice, one needs to determine the renormalization factors $Z_{11}$, $Z_{22}$ and ${\overline {\delta m}}$ non-perturbatively. Each of them can be calculated separately~\cite{Zhang:2017bzy}. Alternatively, the renormalization factor can be calculated as a whole in the regularization-independent momentum subtraction scheme (RI/MOM)~\cite{Martinelli:1994ty}, as was done in the quark case in Refs.~\cite{Stewart:2017tvs,Chen:2017mzz}.

In the RI/MOM scheme, the renormalization factor of the gluon quasi-PDF can be determined by calculating the amputated Green's function of the corresponding operator in a far-offshell gluon state and requiring that the counterterm cancels all loop contributions to this Green's  function at a specific gluon momentum. Schematically, one can write (a detailed investigation of the RI/MOM renormalization will be given elsewhere~\cite{Todo:2018glu})
\begin{align}
&\hspace*{-1.5em}[e^{{\overline{\delta m}}|z|}Z_{11} ^2 Z_3]^{-1}(z n,p^R_z, 1/a, \mu_R)=\non\\
&\hspace*{-2em}\left.\frac{P_{ij}^{ab} \langle 0| T [A^{a,i} (p)  O^{1}(z,0)A^{b,j}(-p) ]|0\rangle|_{amp}}{P_{ij}^{ab} \langle 0|T [A^{a,i} (p)   O^{1}(z,0)A^{b,j} (-p)]|0\rangle_{amp,tree}}\right|_{\tiny\begin{matrix}p^2=-\mu_R^2 \\ \!\!\!\!p_z=p^R_z\end{matrix}},
\end{align}
in the case of $O^1$. Here $A^{a,i}(p)$ denotes the momentum space gluon field with momentum $p$.  $Z_3$ is the gluon field renormalization constant. For $O^{2,3,4}$, $Z_{11} ^2$ should be replaced by  $Z_{22} ^2$, $Z_{11}Z_{12}$, and $Z_{22} ^2$, respectively. $P_{ij}^{ab}$ is a projection operator with color indices $a,b$ and Lorentz indices $i, j$.  A simple example of the projection is $P_{ij}^{ab}= \delta^{ab}g_{\perp,ij}$.
$\mu_R$ and $p_z^R$ are unphysical scales introduced in the RI/MOM scheme~\cite{Stewart:2017tvs,Chen:2017mzz}.

After the non-perturbative renormalization, we can write down the factorization for the renormalized gluon/quark quasi-PDFs:
\begin{align}\label{allfac}
&\tilde g(x ,P_z, p_z^R, \mu_R)=\int_0^1\frac{dy}{y}\Big[C_{gg}\Big(\frac{x}{y}, \frac{\mu_R}{p_z^R},\frac{y P_z}{\mu}, \frac{y P_z}{p_z^R}\Big)g(y, \mu)\non\\
&+C_{gq}\Big(\frac{x}{y}, \frac{\mu_R}{p_z^R},\frac{y P_z}{\mu}, \frac{y P_z}{p_z^R}\Big)q_j(y, \mu)\Big]+\mathcal O\Big(\frac{M^2}{P_z^2}, \frac{\Lambda_{\rm QCD}^2}{P_z^2}\Big),\nonumber\\
&\tilde q_i(x ,P_z, p_z^R, \mu_R)=\int_0^1\frac{dy}{y}\Big[C_{q_i q_j}\Big(\frac{x}{y}, \frac{\mu_R}{p_z^R},\frac{y P_z}{\mu}, \frac{y P_z}{p_z^R}\Big)q_j(y, \mu)\non\\
&+C_{qg}\Big(\frac{x}{y}, \frac{\mu_R}{p_z^R},\frac{y P_z}{\mu}, \frac{y P_z}{p_z^R}\Big)g(y, \mu)\Big]+\mathcal O\Big(\frac{M^2,}{P_z^2}, \frac{\Lambda_{\rm QCD}^2}{P_z^2}\Big),
\end{align}
where the momentum fraction of all parton distributions is within $[0,1]$. A summation of $j$ over all quark/antiquark species is implied. Like the renormalized quasi-PDFs, the hard coefficients $C_{gg}$,  $C_{qg}$, $C_{q_i q_j}$ and $C_{gq}$ also depend on unphysical scales, e.g. $\mu_R, p_z^R$ and the hadron momentum $P_z$. But PDFs extracted from Eq.~(\ref{allfac}) are independent of these scales. For polarized quasi-PDFs, the factorization takes a similar form as Eq.~(\ref{allfac}).
Details of the proof of the above factorization formula using the operator product expansion~\cite{Izubuchi:2018srq}  as well as the perturbative calculation of the hard coefficients  at one-loop accuracy  will be presented in a separate publication~\cite{Todo:2018glu}.

{\bf Conclusion:}
In summary, we have performed a systematic study of the renormalization property of the gluon quasi-PDF defined in LaMET, using an auxiliary adjoint ``heavy quark" Lagrangian. We have shown that all power divergences in the gluon quasi PDF arise from the Wilson line self-energy, and  can be removed by a mass renormalization. We have also identified a set of multiplicatively renormalizable gluon quasi-PDF operators appropriate for lattice simulations. Our findings provide a theoretical basis for directly extracting the gluon PDFs from lattice simulations.


\hspace*{1em}

\begin{acknowledgments}
We thank V. Braun, J.-W. Chen, M. G\"ockeler, Y.-S. Liu, Y.-B. Yang, F. Yuan, Y. Zhao and R.-L. Zhu for helpful discussions.
This work was partially supported by the U.S. Department of Energy Office of Science, Office of Nuclear Physics under Award Number DE-FG02-93ER-40762 and DE-SC0011090 (XJ), and also within the framework of the TMD Topical Collaboration (XJ and AS). XJ acknowledges support from the Alexander von Humboldt Foundation. JHZ and AS are supported by the SFB/TRR-55 grant ``Hadron Physics from Lattice QCD".  This work is also supported in part by National Natural
Science Foundation of China under Grant
 No.11575110, 11655002, 11735010, by Natural Science Foundation of Shanghai under Grant  No.~15DZ2272100.
 \end{acknowledgments}


\begin{thebibliography}{99}

\bibitem{Martinelli:1987zd}
  G.~Martinelli and C.~T.~Sachrajda,
  Phys.\ Lett.\ B {\bf 196}, 184 (1987).
  doi:10.1016/0370-2693(87)90601-0


\bibitem{Martinelli:1988xs}
  G.~Martinelli and C.~T.~Sachrajda,
  Phys.\ Lett.\ B {\bf 217}, 319 (1989).
  doi:10.1016/0370-2693(89)90874-5


\bibitem{Detmold:2001dv}
  W.~Detmold, W.~Melnitchouk and A.~W.~Thomas,
  Eur.\ Phys.\ J.\ direct {\bf 3}, no. 1, 13 (2001)
  doi:10.1007/s1010501c0013
  [hep-lat/0108002].


\bibitem{Dolgov:2002zm}
  D.~Dolgov {\it et al.} [LHPC and TXL Collaborations],
  Phys.\ Rev.\ D {\bf 66}, 034506 (2002)
  doi:10.1103/PhysRevD.66.034506
  [hep-lat/0201021].


\bibitem{Ji:2013dva}
  X.~Ji,
  Phys.\ Rev.\ Lett.\  {\bf 110}, 262002 (2013)
  doi:10.1103/PhysRevLett.110.262002
  [arXiv:1305.1539 [hep-ph]].


\bibitem{Ji:2014gla}
  X.~Ji,
  Sci.\ China Phys.\ Mech.\ Astron.\  {\bf 57}, 1407 (2014)
  doi:10.1007/s11433-014-5492-3
  [arXiv:1404.6680 [hep-ph]].

 
\bibitem{Xiong:2013bka}
  X.~Xiong, X.~Ji, J.~H.~Zhang and Y.~Zhao,
  Phys.\ Rev.\ D {\bf 90}, no. 1, 014051 (2014)
  doi:10.1103/PhysRevD.90.014051
  [arXiv:1310.7471 [hep-ph]].


\bibitem{Ma:2017pxb}
  Y.~Q.~Ma and J.~W.~Qiu,
  Phys.\ Rev.\ Lett.\  {\bf 120}, no. 2, 022003 (2018)
  doi:10.1103/PhysRevLett.120.022003
  [arXiv:1709.03018 [hep-ph]].


\bibitem{Izubuchi:2018srq}
  T.~Izubuchi, X.~Ji, L.~Jin, I.~W.~Stewart and Y.~Zhao,
  Phys.\ Rev.\ D {\bf 98}, no. 5, 056004 (2018)
  [arXiv:1801.03917 [hep-ph]].


\bibitem{Ji:2015qla}
  X.~Ji, A.~Schäfer, X.~Xiong and J.~H.~Zhang,
  Phys.\ Rev.\ D {\bf 92}, 014039 (2015)
  doi:10.1103/PhysRevD.92.014039
  [arXiv:1506.00248 [hep-ph]].


\bibitem{Xiong:2015nua}
  X.~Xiong and J.~H.~Zhang,
  Phys.\ Rev.\ D {\bf 92}, no. 5, 054037 (2015)
  doi:10.1103/PhysRevD.92.054037
  [arXiv:1509.08016 [hep-ph]].

\bibitem{Cichy:2018mum}
  K.~Cichy and M.~Constantinou,
  arXiv:1811.07248 [hep-lat].










\bibitem{Alexandrou:2018pbm}
  C.~Alexandrou, K.~Cichy, M.~Constantinou, K.~Jansen, A.~Scapellato and F.~Steffens,
  arXiv:1803.02685 [hep-lat].


\bibitem{Chen:2018xof}
  J.~W.~Chen, L.~Jin, H.~W.~Lin, Y.~S.~Liu, Y.~B.~Yang, J.~H.~Zhang and Y.~Zhao,
  arXiv:1803.04393 [hep-lat].







\bibitem{Alexandrou:2018eet}
  C.~Alexandrou, K.~Cichy, M.~Constantinou, K.~Jansen, A.~Scapellato and F.~Steffens,
  arXiv:1807.00232 [hep-lat].


\bibitem{Lin:2018qky}
  H.~W.~Lin, J.~W.~Chen, L.~Jin, Y.~S.~Liu, Y.~B.~Yang, J.~H.~Zhang and Y.~Zhao,
  arXiv:1807.07431 [hep-lat].

\bibitem{Liu:2018hxv} 
  Y.~S.~Liu, J.~W.~Chen, L.~Jin, R.~Li, H.~W.~Lin, Y.~B.~Yang, J.~H.~Zhang and Y.~Zhao,
  arXiv:1810.05043 [hep-lat].


\bibitem{Dulat:2015mca}
  S.~Dulat {\it et al.},
  Phys.\ Rev.\ D {\bf 93}, no. 3, 033006 (2016)
  doi:10.1103/PhysRevD.93.033006
  [arXiv:1506.07443 [hep-ph]].


\bibitem{Ball:2017nwa}
  R.~D.~Ball {\it et al.} [NNPDF Collaboration],
  Eur.\ Phys.\ J.\ C {\bf 77}, no. 10, 663 (2017)
  doi:10.1140/epjc/s10052-017-5199-5
  [arXiv:1706.00428 [hep-ph]].


\bibitem{Harland-Lang:2014zoa}
  L.~A.~Harland-Lang, A.~D.~Martin, P.~Motylinski and R.~S.~Thorne,
  Eur.\ Phys.\ J.\ C {\bf 75}, no. 5, 204 (2015)
  doi:10.1140/epjc/s10052-015-3397-6
  [arXiv:1412.3989 [hep-ph]].


\bibitem{Nocera:2014gqa}
  E.~R.~Nocera {\it et al.} [NNPDF Collaboration],
  Nucl.\ Phys.\ B {\bf 887}, 276 (2014)
  doi:10.1016/j.nuclphysb.2014.08.008
  [arXiv:1406.5539 [hep-ph]].


\bibitem{Ethier:2017zbq}
  J.~J.~Ethier, N.~Sato and W.~Melnitchouk,
  Phys.\ Rev.\ Lett.\  {\bf 119}, no. 13, 132001 (2017)
  doi:10.1103/PhysRevLett.119.132001
  [arXiv:1705.05889 [hep-ph]].


\bibitem{Wang:2017qyg}
  W.~Wang, S.~Zhao and R.~Zhu,
  Eur.\ Phys.\ J.\ C {\bf 78}, no. 2, 147 (2018)
  doi:10.1140/epjc/s10052-018-5617-3
  [arXiv:1708.02458 [hep-ph]].


\bibitem{Wang:2017eel}
  W.~Wang and S.~Zhao,
  JHEP {\bf 1805}, 142 (2018)
  doi:10.1007/JHEP05(2018)142
  [arXiv:1712.09247 [hep-ph]].


\bibitem{Fan:2018dxu}
  Z.~Y.~Fan, Y.~B.~Yang, A.~Anthony, H.~W.~Lin and K.~F.~Liu,
  arXiv:1808.02077 [hep-lat].

\bibitem{Ji:2017oey}
  X.~Ji, J.~H.~Zhang and Y.~Zhao,
  Phys.\ Rev.\ Lett.\  {\bf 120}, no. 11, 112001 (2018)
  doi:10.1103/PhysRevLett.120.112001
  [arXiv:1706.08962 [hep-ph]].



\bibitem{Ishikawa:2017faj}
  T.~Ishikawa, Y.~Q.~Ma, J.~W.~Qiu and S.~Yoshida,
  Phys.\ Rev.\ D {\bf 96}, no. 9, 094019 (2017)
  doi:10.1103/PhysRevD.96.094019
  [arXiv:1707.03107 [hep-ph]].

\bibitem{Green:2017xeu}
  J.~Green, K.~Jansen and F.~Steffens,
  Phys.\ Rev.\ Lett.\  {\bf 121}, no. 2, 022004 (2018)
  doi:10.1103/PhysRevLett.121.022004
  [arXiv:1707.07152 [hep-lat]].


























































































\bibitem{Collins:1984xc}
  J.~C.~Collins,
  doi:10.1017/CBO9780511622656



\bibitem{Hatta:2013gta}
  Y.~Hatta, X.~Ji and Y.~Zhao,
  Phys.\ Rev.\ D {\bf 89}, no. 8, 085030 (2014)
  doi:10.1103/PhysRevD.89.085030
  [arXiv:1310.4263 [hep-ph]].




\bibitem{Dorn:1981wa}
  H.~Dorn, D.~Robaschik and E.~Wieczorek,
  Annalen Phys.\  {\bf 40}, 166 (1983).













\bibitem{Bagan:1993zv}
  E.~Bagan and P.~Gosdzinsky,
  Phys.\ Lett.\ B {\bf 320}, 123 (1994)
  doi:10.1016/0370-2693(94)90834-6
  [hep-ph/9305297].




\bibitem{Chen:2017mie}
  J.~W.~Chen, T.~Ishikawa, L.~Jin, H.~W.~Lin, Y.~B.~Yang, J.~H.~Zhang and Y.~Zhao,
  arXiv:1710.01089 [hep-lat].

\bibitem{Zhang:2017bzy}
  J.~H.~Zhang, J.~W.~Chen, X.~Ji, L.~Jin and H.~W.~Lin,
  Phys.\ Rev.\ D {\bf 95}, no. 9, 094514 (2017)
  doi:10.1103/PhysRevD.95.094514
  [arXiv:1702.00008 [hep-lat]].


\bibitem{Maiani:1991az}
  L.~Maiani, G.~Martinelli and C.~T.~Sachrajda,
  Nucl.\ Phys.\ B {\bf 368}, 281 (1992).
  doi:10.1016/0550-3213(92)90528-J

\bibitem{Dorn:1986dt}
  H.~Dorn,
  Fortsch.\ Phys.\  {\bf 34}, 11 (1986).
  doi:10.1002/prop.19860340104


\bibitem{Martinelli:1994ty}
  G.~Martinelli, C.~Pittori, C.~T.~Sachrajda, M.~Testa and A.~Vladikas,
  Nucl.\ Phys.\ B {\bf 445}, 81 (1995)
  doi:10.1016/0550-3213(95)00126-D
  [hep-lat/9411010].

\bibitem{Stewart:2017tvs}
  I.~W.~Stewart and Y.~Zhao,
  Phys.\ Rev.\ D {\bf 97}, no. 5, 054512 (2018)
  doi:10.1103/PhysRevD.97.054512
  [arXiv:1709.04933 [hep-ph]].


\bibitem{Chen:2017mzz}
  J.~W.~Chen, T.~Ishikawa, L.~Jin, H.~W.~Lin, Y.~B.~Yang, J.~H.~Zhang and Y.~Zhao,
  Phys.\ Rev.\ D {\bf 97}, no. 1, 014505 (2018)
  doi:10.1103/PhysRevD.97.014505
  [arXiv:1706.01295 [hep-lat]].


\bibitem{Todo:2018glu}
W. Wang, J.-H. Zhang, and S. Zhao, R.-L. Zhu, to be published.

\end{thebibliography}
\end{document}